# Photodynamic therapy influence on anti-cancer immunity

O.G. Isaeva[*)], V.A. Osipov

*Bogoliubov Laboratory of Theoretical Physics, Joint Institute for Nuclear Research, 141980, Dubna, Russia*

**ABSTRACT**

The system of partial differential equations describing tumor-immune dynamics with angiogenesis taken into account is presented. For spatially homogeneous case, the steady state analysis of the model is carried out. The effects of single photodynamic impact are numerically simulated. In the case of strong immune response we found that the photodynamic therapy (PDT) gives rise to the substantial shrinkage of tumor size which is accompanied by the increase of IL-2 concentration. On the contrary, the photodynamic stimulation of weak immune response is shown to be insufficient to reduce the tumor. These findings indicate the important role of anti-cancer immune response in the long-term tumor control after PDT.

**Kew words** ordinary and partial differential equations, anti-cancer immune response, interleukin-2, angiogenesis, photodynamic therapy

## 1. INTRODUCTION

Photodynamic therapy (PDT) is considered to be one of the most effective methods for treatment of early cancer and palliation of advanced cancer. The US FDA has approved specific forms of PDT for treating esophageal and non-small cell lung cancer, and research shows that the technique can be effective against skin cancer.[1] PDT is based on staining of the tissue by a photosensitizer and subsequent exposing it to low-energy laser radiation of a certain wavelength. At absorption of an optical quantum a long-lived triplet state of photosenstizer molecules is excited. The singlet state of oxygen molecule is generated upon collision of an oxygen molecule in the ground state with a photosensitizer molecule in triplet state. The singlet oxygen is a strong cytotoxic agent which actively destroys cells. It is considered that mainly living cells are damaged under the oxidation of proteins and lipids in the external membrane by the singlet oxygen.[2]

In some experiments the immunostimulating effect of PDT has been found.[3,4] It is shown in Ref. 3 that exposure to a certain dose of 630 nm light next day after the photosensitizer administering does not give prolonged curative effect in severe combined immunodeficient mice whereas mice with normal immune system have no sign of tumor recurrence up to 90 days after the analogous PDT. Moreover, the existing data confirm that the CD8+ T cells are responsible for the long-term tumor control after the PDT. Similar results were obtained in Ref. 4 where it was shown that PDT of murine tumors provides durable inhibition of the growth of untreated lung tumor. The inhibition of tumor growth outside the treatment field is tumor-specific and depends on the presence of CD8+ T cells. These results stimulate our interest to study the role of anti-cancer immune response in curative ability of photodynamic therapy. To this end, in this paper we consider the effects of photodynamic therapy within our mathematical model of anti-cancer immune response suggested in Ref. 5, which is extended here to take into account angiogenesis and spatial inhomogeneity.

---

[*)] Corresponding author. Olga G. Isaeva. Bogoliubov Laboratory of Theoretical Physics, Joint Institute for Nuclear Research, 141980, Dubna, Moscow region, Russia. Tel. +7(49621)63819, Fax: +7(49621)65084, E-mail: issaeva@theor.jinr.ru

## 2. MATHEMATICAL MODEL
### 2.1 Model

Angiogenesis is a physiological process involving the development of new blood vessels from pre-existing ones. Tumor cells induce angiogenesis to obtain necessary nutrients mainly oxygen, which also is the important element in photodynamic reactions. In order to take into account the dynamics of capillary network we add to our model in Ref. 5 the equations for endothelial cells (EC), angiogenesis factor (AF) and normal cells (NC) as well as we modify the model equations to take into account an influence of angiogenesis on anti-cancer immune response.

Optically, biological tissue is a turbid strongly scattering medium. Similarly to Ref. 6 we assume that the intensity of radiation descends with the deepness of penetration as

$$I(z,t) = \begin{cases} I_0 \exp(-(\sigma+\mu_{eff})z), & t \in [T_s, T_f], \\ 0, & t \notin [T_s, T_f], \end{cases}$$

where $\mu_{eff}$ is the effective coefficient of attenuation of radiation, $\sigma$ is the coefficient of absorption of radiation by the photosensitizer molecules in the tissue, $T_s$ and $T_f$ are the time of starting and cessation of irradiation, correspondingly.

In order to take into account spatial inhomogeneity we add the diffusion terms into the model as well as the term describing a motion of endothelial cells in response to AF gradients in the surrounding tissue taken from Ref 7. Since actual geometry of tumors is quite intricate appropriate approximations (e.g. spherical and cylindrical symmetry as well as one-dimensional cases) are used to describe propagation of certain forms of tumors. For our purpose, we simplify the problem by considering the penetration of tumor cells into deeper levels of the tissue as a vertical tumor growth. This could describe a nodular malignant melanoma which has no clinically or histologically evident radial growth phase.[8] Similar approximation is also valid for the description of the growth of subcutaneous (s.c.) experimental tumors where the average values of variables in a plane perpendicular to the direction of motion of the vascular front should be used.

The extended system of partial differential equations for tumor cell population — $T(z,t)$, cytotoxic T lymphocytes (CTL) — $L(z,t)$, interleukin-2 — $I_2(z,t)$, endothelial cells — $E(z,t)$, normal cells — $N(z,t)$, angiogenesis factor - $S(z,t)$ and the fraction of the undamaged vital cellular substratum — $M(z,t)$ is written as (cf. Ref. 5)

$$\frac{\partial T}{\partial t} = D_T \frac{\partial^2 T}{\partial z^2} - aT \ln \frac{bT}{aE} - cTL - r_1(1-M)T, \tag{1}$$

$$\frac{\partial L}{\partial t} = D_L \frac{\partial^2 L}{\partial z^2} + \frac{dE}{E+E_0} + eLI_2 - fL - r_2(1-M)L, \tag{2}$$

$$\frac{\partial I_2}{\partial t} = \frac{gT}{(T+l)(\alpha S+1)} - jLI_2 - kTI_2 - m_1 I_2, \tag{3}$$

$$\frac{\partial E}{\partial t} = D_E \frac{\partial^2 E}{\partial z^2} - \chi_0 \frac{\partial}{\partial z}\left(E \frac{\partial S}{\partial z}\right) - f_1 E + qSE - r_3(1-M)E, \tag{4}$$

$$\frac{\partial N}{\partial t} = D_N \frac{\partial^2 N}{\partial z^2} + \frac{g_1 EN_0}{E+E_0} - f_2 N - k_1 TN - r_4(1-M)N, \tag{5}$$

$$\frac{\partial S}{\partial t} = D_S \frac{\partial^2 S}{\partial z^2} + \frac{p_4 T^2}{T^2+\tau_s^2} + \frac{g_2 E_0 N_0}{N+N_0} - j_1 SE - m_2 S, \tag{6}$$

$$\frac{\partial M}{\partial t} = \beta(1-M) - \frac{p_{ox} I(z,t)EM}{(E+E_0)(M+K_M)}. \tag{7}$$

Equations (1)—(6) describe the tumor-immune dynamics with taking account of vascularity. Based on Ref. 9, the carrying capacity of tumor cell population is taken to be proportional to the density of the endothelial cells. When the exist-ing capillary network becomes insufficient to supply tumor cells with nutrients, they start to produce AF, for example the vascular endothelial growth factor (VEGF). In the extended model the production of AF by tumor cells is described by the second term in (6), similarly to Ref. 10. The suppression of the immune response caused by growing concentration of AF is also taken into account.[10,11] It is supposed that the influx of CTL into the tissue depends on the amount of the EC.

Equation (7) for the fraction of undamaged substratum is formulated based on the kinetic models of photodynamic reactions proposed in Refs. 6,12. In Ref. 6 the steady state concentration of the singlet oxygen is shown to be proportional to the light intensity and the photosensitizer concentration. At low concentrations of the oxygen the steady state concentration of the singlet oxygen is nearly linear in the oxygen concentration, at higher concentrations of the oxygen the concentration of singlet oxygen turns out to be independent on the concentration of oxygen. In the model proposed in Ref. 12 physical and chemical quenching of the singlet oxygen is considered. Three mechanisms of physical quenching are possible: the transfer of energy from the singlet oxygen to lower triplet levels of the molecules of quenching agent; the transfer of energy to the oscillating sublevels of the quenching agent molecules; the formation of complexes with the charge transfer between the singlet oxygen and molecules of the quenching agent.[12] Chemical quenching is the oxidization of the quenching agent. In contrast to Ref. 6 photochemical reactions are considered in Ref. 12 in the absence of hypoxia. Within our study the supply tissue with oxygen depends on amount of the endothelial cells, which are distributed nonuniformly in tumor site. Therefore the rate of the substratum oxidization (proportional to the steady state concentration of the singlet oxygen) is described by the term $p_{ox}I(z,t)EM/(E+E_0)(M+K_M)$. Similarly to Ref.12 the first term in (7) is the repair rate of damaged cellular substratum. The terms describing the death rate of cells after PDT in equations (1), (2), (4), (5) are also taken from Ref.12.

**2.2 Model parameters**

The dynamics of disease is very sensitive to the parameters in equations (1)—(7). In this paper, we restrict our consideration to a study of the photodynamic influence on development of the s.c. experimental tumor in mice. In this case, it is appropriate to use the parameter set M2 from Ref. 5. For convenience let us divide the parameters into two groups: system and therapeutic parameters. The system parameters are presented in the Table 1. Some of them were estimated from experimental data (see Ref. 5 for details). The values of $m_1$ and $m_2$ are calculated basing on the data for half lifetimes of cytokines (see, e.g., Ref. 13). The values of $p_4$ and $\tau_s$ are chosen in accordance with Ref. 10. The diffusion coefficients for cellular populations and AF are taken from Ref. 14. The chemotactic parameter of endothelial cells is obtained using data presented in Refs. 7,11,15. For the rest of system parameters we choose values most appropriate to our model. Current medical literature and sensitivity analysis allow us to conclude that the corresponding interactions are of importance in the description of tumor-immune dynamics.

The parameters, characterizing the photodynamic influence are presented in the Table 2. First of all, let us consider the coefficient of attenuation of radiation $\mu_{eff}$ and the coefficient of absorption of radiation by the photosensitizer molecules in the tissue $\sigma$. The value of $\mu_{eff}$ is taken from Ref. 16. The coefficient $\sigma$ is equal to $\sigma_0[PS]^S(t)$, where $\sigma_0$ is a transition cross section from the ground state to the first excited state of photosensitizer, $[PS]^S$ is a concentration of the photosensitizer molecules in the ground state. The transition cross section is estimated based on the data presented in Ref. 6. It is equal approximately to $6,3 \cdot 10^{-19}$ cm$^2$. In our consideration, the concentration $[PS]^S(t)$ is supposed to be in the steady state and does not change during exposure to laser (about 0.5 h). First, as is estimated in Ref. 6 the concentration of singlet oxygen riches its steady state in about 3 μs. Hence, the time for quenching of exited triplet state of photosensitizer upon collision with the molecules of oxygen is smaller than the time of irradiation. Second, the duration of exposure is smaller in comparison with the half lifetime of photosensitizer in the tissue. In order to estimate the photosensitizer concentration we assume that its entire dose is accumulated in the tumor site. In the framework of one-dimensional problem, we consider that tumor grows in the cylindrical volume with 5 mm in diameter and 2 mm in thickness. Thus, the concentration of photosensitizer introduced with dose 5 mg/kg in this volume is estimated to be $2.5 \times 10^{-3}$ g/cm$^3$.

The parameter characterizing the rate of oxidization of substratum is defined by $p_{ox} = \sigma_0[PS]^S\varphi/(h\nu[M]_0)$,[6,12] where $\varphi$ is the quantum yield of singlet triplet interconversion of the photosensitizer molecule, $[M]_0$ is the concentration of sensitive substratum, $h\nu$ is the energy of laser quantum. Assuming the quantum yield $\varphi$ equal to 1, we obtain for $p_{ox}$ the value 0.114 cm$^2 \cdot$J$^{-1}$. The parameter $K_M$ is calculated by using the relation $(t_q k_{ox}[M]_0)^{-1}$.[12] The half lifetime of the singlet state of oxygen molecule $t_q$ is taken to be 1 μs.[2] The average rate of oxidization of proteins and lipids $k_{ox} = 5 \cdot 10^7$ M$^{-1}$·c$^{-1}$ is estimated by using the data for the oxidization constants of proteins and lipids in Ref. 2.

Table 1. Parameter set

| Parameter | Units | Description | Value | Source |
|---|---|---|---|---|
| $a$ | day$^{-1}$ | Tumor growth rate | 0,22 | [5] |
| $b/E_0$ | cell$^{-1}$·mm·day$^{-1}$ | $aE(t)/b$ is carrying capacity of the tumor cell population | 1.68×10$^{-7}$ | [5] |
| $c$ | mm·cell$^{-1}$·day$^{-1}$ | Rate of inactivation of tumor cells by CTL | 5.6×10$^{-7}$ | [5] |
| $d$ | cell·mm$^{1}$·day.$^{-1}$ | Influx rate of CTL | 7.9×10$^4$ | [5] |
| $e$ | mm·units$^{-1}$ day$^{-1}$ | Rate of CTL proliferation in response to IL-2 | 2.24×10$^{-8}$ | [5] |
| $f$ | day$^{-1}$ | CTL death rate | 0,33 | [5] |
| $g$ | units·day$^{-1}$ | Antigen presentation | 6.25×10$^6$ | [5] |
| $j$ | mm·cell$^{-1}$·day$^{-1}$ | Rate of consumption of IL-2 molecules by CTL | 1.32×10$^{-7}$ | [5] |
| $k$ | mm·cell$^{-1}$·day$^{-1}$ | Inactivation of IL-2 molecules by prostaglandins | 1.1×10$^{-6}$ | [5] |
| $l$ | cell·mm$^{-1}$ | Half-saturation constant | 5×10$^4$ | [5] |
| α | units$^{-1}$·cm | 1/α is the amount of AF for which the antigen presentation descend 2 fold | 2.5×10$^{-10}$ | |
| $m_1$ | day$^{-1}$ | IL-2 elimination rate | 0.02 | [13] |
| $m_2$ | day$^{-1}$ | AF elimination rate | 0.007 | [13] |
| $f_1$ | day$^{-1}$ | Endothelial cells loss rate | 0.023 | |
| $f_2$ | day$^{-1}$ | Normal cells loss rate | 0.03 | |
| $q$ | units$^{-1}$·cm·day$^{-1}$ | Rate of proliferation of endothelial cells in response to AF | 1.32×10$^{-10}$ | |
| $g_1$ | day$^{-1}$ | Normal cells influx | 0.06 | |
| $g_2$ | unit·cell$^{-1}$·day$^{-1}$ | AF production by normal cells | 3168 | |
| $k_1$ | mm·cell$^{-1}$·day.$^{-1}$ | Loss rate of normal cells in the presence of tumor cells | 5.8×10$^{-8}$ | |
| $p_4$ | units·cm·day$^{-1}$ | AF production by tumor cells | 5.3×10$^7$ | [10] |
| $τ_S$ | cell·mm$^{-1}$ | Half saturation constant for AF production | 5×10$^5$ | [10] |
| $j_1$ | mm·cell$^{-1}$·day$^{-1}$ | Rate of consumption of AF by endothelial cells | 6.3×10$^{-5}$ | |
| $D_T$ | cm$^2$·s$^{-1}$ | Diffusion coefficient of tumor cells | 2.5×10$^{-10}$ | [14] |
| $D_L$ | cm$^2$·s$^{-1}$ | Diffusion coefficient of CTL | 2.5×10$^{-10}$ | [14] |
| $D_E$ | cm$^2$·s$^{-1}$ | Diffusion coefficient of endothelial cells | 2.5×10$^{-11}$ | [14] |
| $D_N$ | cm$^2$·s$^{-1}$ | Diffusion coefficient of normal cells | 2.5×10$^{-10}$ | [14] |
| $D_S$ | cm$^2$·s$^{-1}$ | Diffusion coefficient of AF | 7.5×10$^{-7}$ | [14] |
| $χ_0$ | cm$^3$·s$^{-1}$·units$^{-1}$ | Chemotactic coefficient of endothelial cells | 7.0×10$^{-18}$ | [7,11,15] |

Table 2. Parameters of photodynamic influence

| Parameter | Units | Description | Value | Source |
|---|---|---|---|---|
| $\sigma$ | cm$^{-1}$ | Coefficient of absorption of radiation | 1.0 | [6] |
| $\mu_{eff}$ | cm$^{-1}$ | Effective coefficient of attenuation of radiation | 5.0 | [16] |
| $p_{ox}$ | cm$^2\cdot$J$^{-1}$ | Rate of oxidization of substratum | 0.114 | [2,6] |
| $I_0$ | mW$\cdot$cm$^{-2}$ | Laser light intensity | 75.0 | [4] |
| $K_M$ | | Half saturation constant | 0.5 | [2,12] |
| $\beta$ | day$^{-1}$ | Rate of reparation of substratum | 3.3 | |
| $r_1$ | day$^{-1}$ | Tumor cell killing | 1.32 | |
| $r_2$ | day$^{-1}$ | CTL cell killing | 1.0 | |
| $r_3$ | day$^{-1}$ | Endothelial cells killing | 1.0 | |
| $r_4$ | day$^{-1}$ | Normal cells killing | 1.0 | |

## 3. STEADY STATE ANALYSIS OF THE SPATIALLY HOMOGENEOUS SYSTEM

In order to simplify analysis and numerical simulations we introduce dimensionless variables: $T' = T/T_0$, $L' = L/L_0$, $I'_2 = I_2/I_{20}$, $E' = E/E_0$, $N' = N/N_0$, $S' = S/S_0$, where $T_0 = 1.3\times10^6$ cell$\cdot$mm$^{-1}$, $L_0 = 5\times10^5$ cell$\cdot$mm$^{-1}$, $I_{20} = 10^7$ unit$\cdot$mm$^{-1}$, $E_0 = 2.5\times10^3$ cell$\cdot$mm$^{-1}$, $N_0 = 5\times10^6$ cell$\cdot$mm$^{-1}$, $S_0 = 10^9$ unit$\cdot$cm$^{-1}$.

We perform the steady state analysis of the autonomous system of ordinary differential equations (ODE) that describes the underlying spatially homogeneous kinetics of (1)—(6). This ODE system is obtained assuming the diffusion coefficients to be equal to infinity. Actually, the dimensionless diffusion coefficients $D' = D\tau/L^2$ go up to infinity while the length of the region under consideration, $L$, tends to zero, i.e. the region descends to the point. Thus, the dispersed system is reduced to the point model.

The linear stability analysis shows that the tumor free steady state $(0, L_0^*, 0, E_0^*, N_0^*, S_0^*)$ is always unstable. This means that even one malignant cell can develop tumor cell population and the spontaneous full tumor regression is not possible.

A bifurcation diagram for the model parameter characterizing the production of AF by tumor cells is presented in Fig. 1. As is seen the system has two bifurcation points. Therefore one can distinguish three main dynamical regimes. First of all, at low AF production rate ($p_4 < p_{4min}$) the tumor grows slowly and reaches the small size $\bar{T}_1$ where the equilibrium between the tumor and immune system is established. This dynamics corresponds to the dormant growth of the tumor. The enhancement of AF production rate shifts the system into the region ($p_{4min} < p_4 < p_{4max}$), where two stable fixed points (corresponding to the tumor sizes $\bar{T}_1$ and $\bar{T}_3$) and one unstable fixed points ($\bar{T}_2$) exist. In this case, depending on the initial conditions two scenario of the tumor development are possible: the tumor remission and uncontrolled growth. For high AF production rate ($p_4 > p_{4max}$) the stable fixed point corresponding to the small size of tumor $\bar{T}_1$ and unstable fixed point ($\bar{T}_2$) disappear. Hence, the progressive growth of tumor suppressing the immune response takes place.

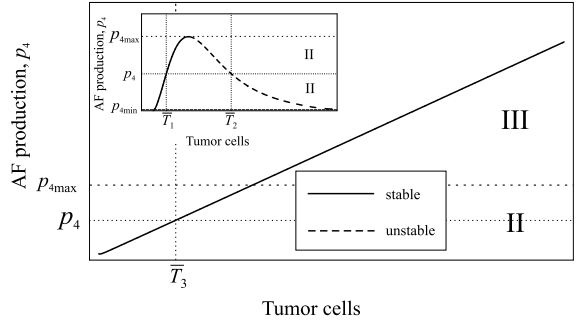

Fig. 1. Bifurcation diagram for parameter, $p_4$, characterizing the AF production. The inset highlights the region of low values of $p_4$.

## 4. NUMERICAL EXPERIMENTS

We simulate the single exposure (during 0.5 h) of previously photosensitized tissue to laser light with intensity 75 mW/cm$^2$ and wavelength 630 nm. The chosen parameters in Table 1 provide for two possible dynamical regimes: regression of tumor to the small size and progressive growth.

In order to perform numerical simulations we assume the following initial conditions

$$T(z,0) = \begin{cases} T, & 0 \le z \le 0.4 \\ T\exp(-60(z-0.4)^2), & 0.4 < z \le 1 \end{cases}, \quad L(z,0) = L\exp(-20(z-0.5)^2),$$

$$I_2(z,0) = I_2\exp(-15(z-0.5)^2), \quad E(z,0) = E\exp(-30(z-0.5)^2),$$

$$N(z,0) = \begin{cases} N\exp(-60(z-0.4)^2), & 0 \le z \le 0.4 \\ N, & 0.4 < z \le 1 \end{cases}, \quad S(z,0) = S\exp(-20z^2), \quad M(z,0) = 1.$$

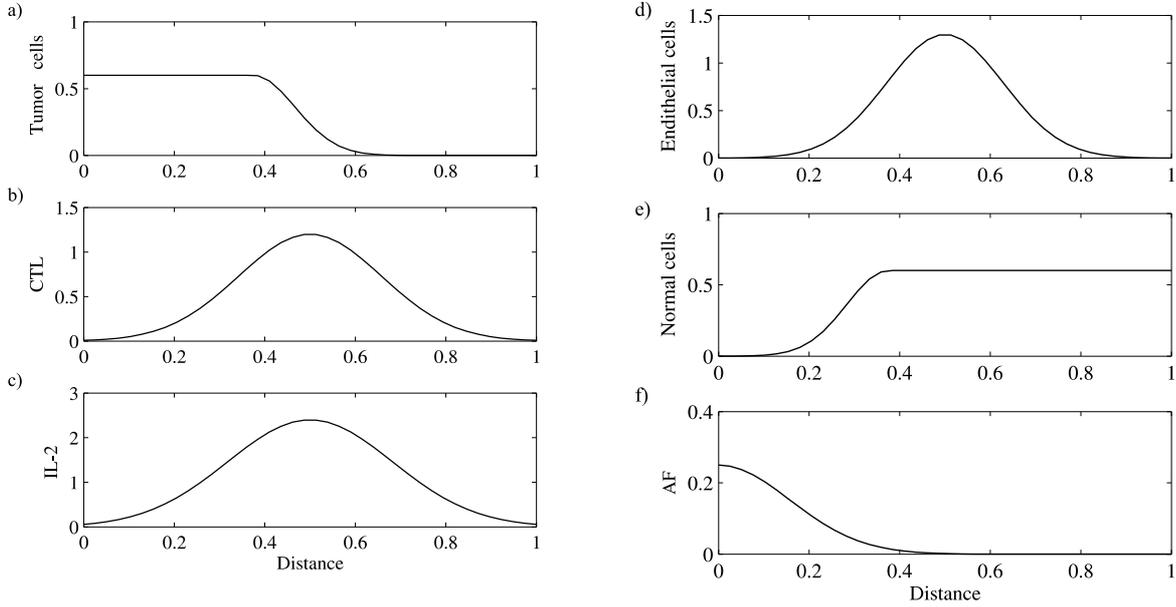

Fig. 2. Initial profiles for densities of cellular populations: a) tumor cells, b) CTL, c) IL-2, d) endothelial cells, e) normal cells and f) AF. All values are given in reduced units

The initial profiles are presented in Fig. 2. We suppose that primarily there is the small amount of endothelial cells and CTL in the inoculated experimental tumor, particularly near the external wall. Let us consider the photodynamic influ-ence for two cases of initial densities of the endothelial cells: $E_1\exp(-30(z-0.5)^2)$ and $E_2\exp(-30(z-0.5)^2)$ where $E_2 > E_1$. Thus, we reflect the influence of the vascular level on the tumor-immune dynamics. It is expected that in the case of high vascularity the immune response can be weakened. Therefore, the cases of high and low vascularities are considered as weak and strong immune responses, correspondingly. Zero-flux boundary conditions are imposed on the variables $T$, $L$, $E$, $N$, $S$, which are equivalent to

$$\vec{n} \cdot \nabla T = \vec{n} \cdot \nabla L = \vec{n} \cdot \nabla E = \vec{n} \cdot \nabla N = \vec{n} \cdot \nabla S = 0.$$

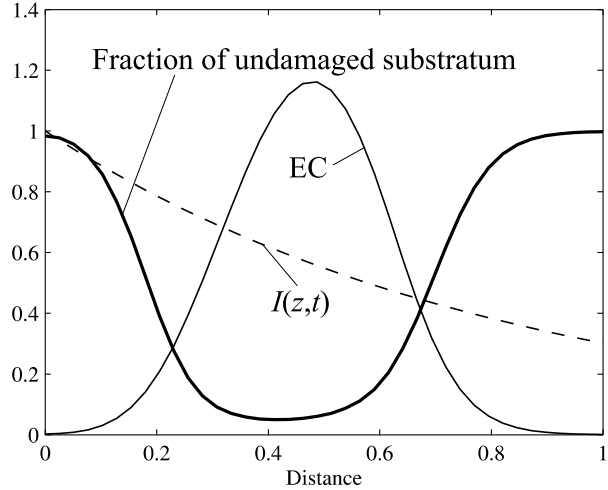

Fig. 3. The spatial distribution of the oxidized substratum, laser light intensity and endothelial cells. All values are given in reduced units

In Fig. 3 the distribution of the oxidized substratum, laser light intensity and density of endothelial cells are shown. As is seen the largest damage occurs in the region where the population of endothelial cells and intensity are high enough. These results show that the photodynamic effectiveness depends on the oxygen saturation of the tissue. In Figs. 4 and 5 the distribution of the densities of cellular populations on 100th day of consideration for strong and weak immune response is presented. It is seen that in the case of strong immune response the tumor dynamics changes markedly after single PDT impact, namely the tumor decreases to a small size. At the same time both CTL and IL-2 concentration after PDT are higher than in the case without PDT. Thus, the small damage of tumor cell population by PDT causes the stimulation of immune response sufficient to control the tumor cells. In contrast, in the system with weak immune response the tumor proceeds with growing up to the maximum size despite the slight elevation of CTL and IL-2 concentrations.

## 5. CONCLUSION

The extended model of tumor-immune dynamics with angiogenesis taken into account is considered. The system consists of six partial differential equations for tumor cells, cytotoxic T cells, interleukin-2, endothelial and normal cells as well as angiogenesis factor. The steady state analysis of underlying homogeneous system shows three main regimes of tumor immune interactions depending on the AF production rate. At low AF production the tumor is handled by the immune response and grows to the small stable size. This behaviour corresponds to the case of dormant tumor. At medium AF production the initial sizes of tumor cells populations and CTL have crucial influence on the disease outcome. Two regimes are possible: a progressive increase of tumor population to the highest possible size and the tumor remission. For high AF production the regime of tumor remission becomes impossible. The tumor is able to evade the immune surveillance. The influence of single photodynamic therapy impact on the system dynamics is considered. It is shown that the regime of tumor remission can be achieved after the PDT in the case of strong immune response. Besides, for both conditions of the immune response the increase of the interleukin-2 concentration is found in comparison with the case without PDT. However, the PDT stimulation of weakened immune response is shown to be insufficient to provoke the tumor remission. This allows us to conclude that the curative ability of PDT is markedly governed by the immune system condition.

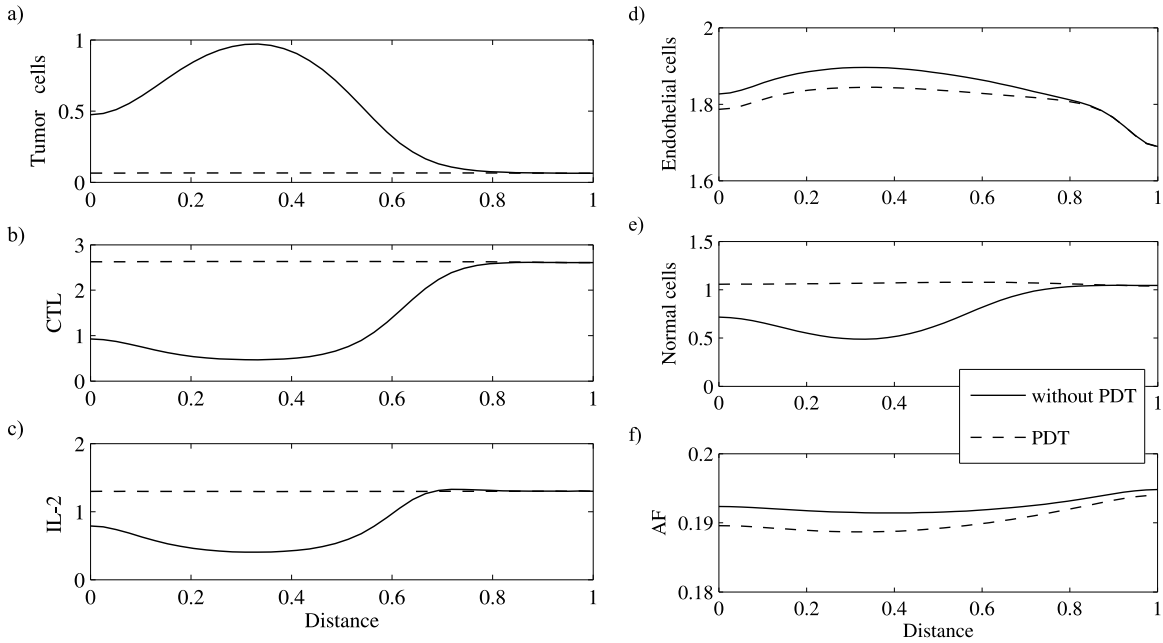

Fig. 4. Spatial distributions of: a) tumor cells, b) CTL, c) IL-2, d) endothelial cells, e) normal cells and f) AF densities on 100th day for strong immune response. All values are given in reduced units.

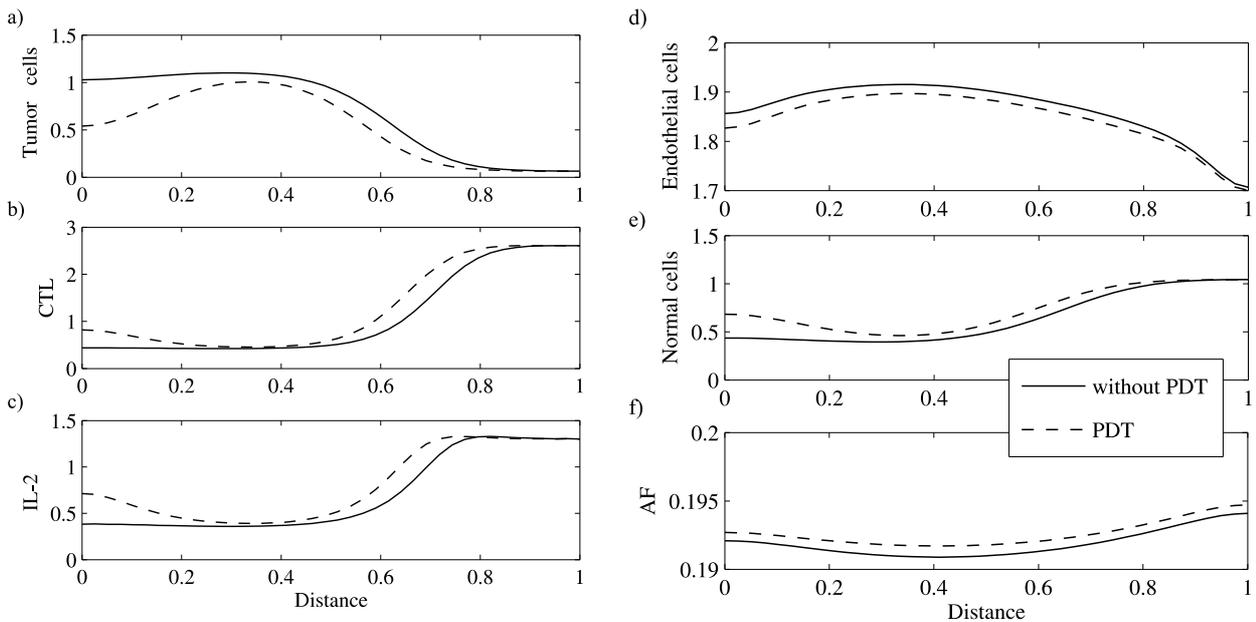

Fig. 5. Spatial distributions of: a) tumor cells, b) CTL, c) IL-2, d) endothelial cells, e) normal cells and f) AF densities on 100th day for weak immune response. All values are given in reduced units.